\documentclass[
reprint,
superscriptaddress,
amsmath,amssymb,
aps,
]{revtex4-2}

\usepackage{graphicx}   
\usepackage{comment}
\usepackage{dcolumn}    
\usepackage{bm}         
\usepackage{hyperref}   

\usepackage{color}

\begin{document}

\title{Conductance asymmetry in proximitized magnetic\\ topological insulator junctions with Majorana modes}

\author{Daniele Di Miceli}
\affiliation{Institute for Cross-Disciplinary Physics and Complex Systems IFISC (CSIC-UIB), E-07122 Palma, Spain}
\affiliation{Department of Physics and Materials Science, University of Luxembourg, 1511 Luxembourg, Luxembourg}

\author{Eduárd Zsurka}
\affiliation{Department of Physics and Materials Science, University of Luxembourg, 1511 Luxembourg, Luxembourg}
\affiliation{Peter Grünberg Institute (PGI-9), Forschungszentrum Jülich, 52425 Jülich, Germany}
\affiliation{JARA-Fundamentals of Future Information Technology, Jülich-Aachen Research Alliance, Forschungszentrum Jülich and RWTH Aachen University, Germany}

\author{Julian Legendre}
\affiliation{Department of Physics and Materials Science, University of Luxembourg, 1511 Luxembourg, Luxembourg}

\author{Kristof Moors}
\affiliation{Peter Grünberg Institute (PGI-9), Forschungszentrum Jülich, 52425 Jülich, Germany}
\affiliation{JARA-Fundamentals of Future Information Technology, Jülich-Aachen Research Alliance, Forschungszentrum Jülich and RWTH Aachen University, Germany}

\author{Thomas L. Schmidt}
\affiliation{Department of Physics and Materials Science, University of Luxembourg, 1511 Luxembourg, Luxembourg}
\affiliation{School of Chemical and Physical Sciences, Victoria University of Wellington, P.O. Box 600, Wellington 6140, New Zealand}

\author{Llorenç Serra}
\affiliation{Institute for Cross-Disciplinary Physics and Complex Systems IFISC (CSIC-UIB), E-07122 Palma, Spain}
\affiliation{Department of Physics, University of the Balearic Islands, E-07122 Palma, Spain}

\begin{abstract}
We theoretically discuss electronic transport via Majorana states in magnetic topological insulator-superconductor
junctions with an asymmetric split of the applied bias voltage.
We study normal-superconductor-normal (NSN) junctions made of narrow (wire-like) or wide (film-like) magnetic topological insulator slabs with a central proximitized superconducting sector.
The occurrence of charge non-conserving Andreev processes entails a nonzero conductance related to an electric current flowing to ground from the proximitized sector of the NSN junction.
We show that topologically-protected Majorana modes require an antisymmetry of this conductance with respect to the point of equally split bias voltage across the junction.
\end{abstract}

\maketitle

\section{Introduction
\label{sec:Introduction}}

Majorana modes in solid-state physics are zero-energy quasiparticle excitations with the unusual property of being their own antiparticles \cite{Majorana, Majorana_returns}, which emerge in 1D and 2D topological superconductors (TSCs) near boundaries and vortices \cite{Majorana_solid_state}.
These fascinating states can be distinguished into Majorana chiral propagating states (MCPS) in two-dimensional superconducting phases \cite{Majoranas_TSC, TSC_2D_and_3D, Topological_Ins_Sc}, and zero-energy Majorana bound states (MBS) in spinless $p$-wave superconducting wires \cite{Kitaev_chain}.
The former are dispersive modes analogous to quantum anomalous Hall and quantum spin Hall edge states in superconducting materials \cite{TSC_2D_and_3D, Topological_Ins_Sc}, while the latter are localized in-gap modes emerging at the ends of gapped phases of 1D topological superconducting wires.
Achieving topological superconductivity is a crucial step toward the realization of non-Abelian braiding statistics and fault-tolerant quantum computing \cite{Non-abelyan_anyons, QC_anyons, Braiding_Chiral_Vortices}.
Magnetic topological insulators \cite{magnetic_TIs}, i.e., 3D topological insulators (TIs) with topological surface states and ferromagnetic ordering, are promising candidates for the realization of such robust platforms for quantum computation, since in presence of proximity coupling to an ordinary $s$-wave superconductor they are expected to realize different TSCs with either propagating or localized Majorana modes \cite{Chiral_TSC, Chiral_TSC_Half-Integer_Plateau, QAH_Majorana_Platform, Quasi-1D-QAH-Majorana}.

Despite the growing interest in proximitized MTIs \cite{Progress_MTIs}, the experimental detection of Majorana modes is still inconclusive \cite{Retracted_MCPS, Editorial_Retraction, Absence_Evidence, Retracted_MBS, Retraction-Note_MBS}.
In this paper, we highlight a characteristic feature of Majorana states that can be used in their detection.
Through theoretical analysis and numerical simulations, we find that both types of Majorana states lead to a peculiar transport signature in NSN junctions between normal (N) and proximitized (S) magnetic topological insulators, when the bias between the two N sections is split asymmetrically with respect to the central S lead.
Without Majorana modes or trivial Andreev bound states (ABSs), which may be found in non-topological 1D superconductors \cite{ABS_review, ABS_introduction}, the electric currents flowing through the N leads are equal and opposite, independently of how the bias is split between left and right leads.
In the presence of Andreev processes, instead, the currents in the two N leads can be different, depending on the fraction of bias applied to each side of the junction.
When the electric currents in the N leads have different intensities, charge conservation requires the existence of a third current going to ground from the superconductor, defining a nonzero differential conductance.

We show in this work that in presence of nontrivial MBSs or MCPSs, the NSN conductance of an MTI slab must be \emph{antisymmetric} with respect to the splitting of the bias. This reflects the existence of identical scattering amplitudes at the two interfaces of the junction.
Observing how the total conductance varies with the bias splitting provides a selective criterion, although not absolutely conclusive, to rule out electric signals coming from trivial ABS in the proximitized MTI slabs, and constitutes a novel alternative approach for transport measurements in NSN junctions.
Monitoring the conductance asymmetry with a continuous change of the bias split is more selective than just observing zero-bias conductance peaks when the full bias drop is applied in turn to each side of the junction.
Furthermore, the proposed symmetry analysis can be useful to discriminate between the two different types of Majorana excitations which can be found in MTIs, and provide an additional control parameter, i.e., the bias split, while maintaining the correlation between the transport behaviour on the two interfaces of the junction.
Similar criteria to detect MBSs on the ends of proximitized semiconducting wires have been discussed in recent works, with a focus on multi-terminal transport measurements \cite{Transport-1, Transport-2, Transport-3, Transport-4, Protocol} and noise correlations \cite{Noise-1, Noise-2}.
Conductance matrix symmetries
of particle-hole type
have  also been investigated \cite{Conductance-Symmetries-1, Conductance-Symmetries-2, Conductance-Symmetries-3}, discussing the
inversion of a common potential acting on all normal leads attached to the superconducting sector.
By contrast, the symmetry
discussed in our work corresponds
to the role of a bias between the two normal leads of our three-terminal setup.

The paper is structured as follows. In Sec.~\ref{sec:Hamiltonian} we discuss
the model Hamiltonian and the topological states in finite-size MTI slabs.
In Sec.~\ref{sec:Conductance} we compute the electric conductance in the NSN junction and discuss its symmetry properties with bias splitting.
In Sec.~\ref{sec:Numerical} we show some numerical results supporting our conclusions.
Sec.~\ref{sec:Conclusion} concludes the manuscript.

\section{Model Hamiltonian}
\label{sec:Hamiltonian}

To start, we consider the Hamiltonian of a 3D TI in presence of ferromagnetic ordering.
In the basis
$\phi_{k\sigma}^\tau = ( c_{k \uparrow}^{+}, c_{k \uparrow}^{-}, c_{k \downarrow}^{+}, c_{k \downarrow}^{-})^T$,
where $c_{k \sigma}^{\tau} \equiv c_{k \sigma}^{\tau}(y,z)$ annihilates an electron with longitudinal wave number $k \equiv k_x$, spin $\sigma=\uparrow, \downarrow$ and orbital index $\tau=\pm$,
the effective 3D Hamiltonian for magnetic TIs takes the following form \cite{Zhang_TI_Model, Zhang_Hamiltonian}
\begin{gather}\label{eq:MTI_Hamiltonian}
    \mathcal{H}_0 (\mathbf{k}) = \epsilon(\mathbf{k}) + M(\mathbf{k}) \tau_z + A(\mathbf{k}) \tau_x + \Lambda \sigma_z \,,
\end{gather}
where
\begin{equation}
\begin{split}
    \epsilon(\mathbf{k}) & = \mu - C_\perp \left( k_x^2+\hat{k}_y^2 \right) - C_z \hat{k}_z^2 \,, \\
    M(\mathbf{k}) & = M_0 - M_\perp \left( k_x^2+\hat{k}_y^2 \right) - M_z \hat{k}_z^2 \,, \\
    A(\mathbf{k}) & = A_\perp \left( k_x \sigma_x + \sigma_y \hat{k}_y \right) + A_z \sigma_z\hat{k}_z \,.
\end{split}
\end{equation}
Here, $\mathbf{k}=(k_x,\hat{k}_y,\hat{k}_z)$ and the transverse momentum operators are given by $\hat{k}_y=-i \hbar \partial_y$ and $\hat{k}_z=-i \hbar \partial_z$.
The Pauli matrices $\sigma_i$ and $\tau_i$ ($i \in \{x,y,z\}$) act on the spin and orbital subspaces, respectively, the magnetization along $z$ is represented by the Zeeman term $\Lambda  \sigma_z$, and $\mu$ is the chemical potential.
This Hamiltonian is suitable for describing TIs such as Bi$_2$Se$_3$, Bi$_2$Te$_3$ and Sb$_2$Te$_3$ through a proper choice of parameters \cite{Zhang_TI_Model, Zhang_Hamiltonian}.
In our simulations, we used the values given in Ref.~\cite{Magnetotransport_signatures_TIs} for a topological insulator where the asymmetry between the conduction and valence bands as well as the anisotropy of the Dirac cones have been neglected.
When placed in proximity to a superconductor, the system can be described by the Bogoliubov-de Gennes (BdG) Hamiltonian \cite{BdG_Theory}
\begin{equation}\label{eq:BdG_Hamiltonian}
    \mathcal{H}_{\text{BdG}}(\mathbf{k})     =
    \begin{pmatrix}
        \mathcal{H}_0(\mathbf{k}) & \Delta^\star \\
        \Delta & -\sigma_y \mathcal{H}_0^\star(-\mathbf{k}) \sigma_y
    \end{pmatrix}
    \,,
\end{equation}
expressed in the basis \cite{BdG_basis}
\begin{equation}\label{eq:BdG_wavefunction}
    \Phi_{k \sigma}^\tau =
    \begin{pmatrix}
        c_{k \uparrow}^{+}, c_{k \uparrow}^{-}, c_{k \downarrow}^{+}, c_{k \downarrow}^{-},
        -c_{-k \downarrow}^{+ \dagger}, -c_{-k \downarrow}^{- \dagger}, c_{-k \uparrow}^{+ \dagger}, c_{-k \uparrow}^{- \dagger}
    \end{pmatrix}^T
    \,,
\end{equation}
where we assumed a local $s$-wave pairing with amplitude $\Delta \equiv \Delta(y,z)$ induced by proximity.
In the following, we fix the thickness of the slab to $d=4$ nm and consider a wire-like geometry with width $L_y=20$ nm and a film-like one with $L_y=160$ nm.
For $d=4$ nm, the surface states on opposite sides of the MTI slab are coupled \cite{Crossover_2D-Limit, Finite-size_Gap_Bi2Se3}, and a finite size gap opens up in the energy spectrum.
The magnetization can induce a gap inversion, yielding nontrivial topological states.
We will assume a constant pairing field along $y$, and model the proximity coupling on the upper surface of the magnetic TI by
\begin{equation}
    \Delta(y,z) = \Delta\, \Theta(z - d/2) \,.
\end{equation}
where $\Theta$ is the Heaviside step function. An asymmetric pairing on the top and bottom surfaces is indeed required to achieve topological superconductivity in the MTI slab \cite{Chiral_TSC_Half-Integer_Plateau, Quasi-1D-QAH-Majorana}.

All the numerical results below are obtained with $\Delta=5$ meV for the wire and $\Delta=10$ meV for the film geometry. These values are unrealistically large compared to experiments, but they are convenient for numerical simulations and qualitatively similar results can be obtained for smaller pairings and rescaled systems.
Indeed, in a quasi-1D superconducting wire, the decay length (along the longitudinal direction $x$) of Majorana end states is inversely proportional to the pairing potential $\xi \propto 1/|\Delta|$.
This means that, in order to guarantee well-separated MBSs, a smaller pairing can be compensated by a greater length $L_x$, as long as the ratio $\xi/L_x$ is unchanged.
Similarly, in the effective 2D superconductor, the edge modes localization length $l_c$ (along the transverse direction $y$) scales with the inverse of the pairing amplitude: a smaller gap requires the thin film to be wider to maintain a constant ratio $\l_c/L_y$  and ensure decoupled edge modes.
Therefore, a larger pairing $\Delta$ allows us to reduce the computational effort by using smaller systems and, at the same time, gives us the opportunity to enhance the energy gap for MBSs and increase the width of the region with MCPSs.
A similar scaling has already been proposed in graphene \cite{Scalable_Graphene}.

\begin{figure}
    \centering
    \includegraphics[width=\linewidth]{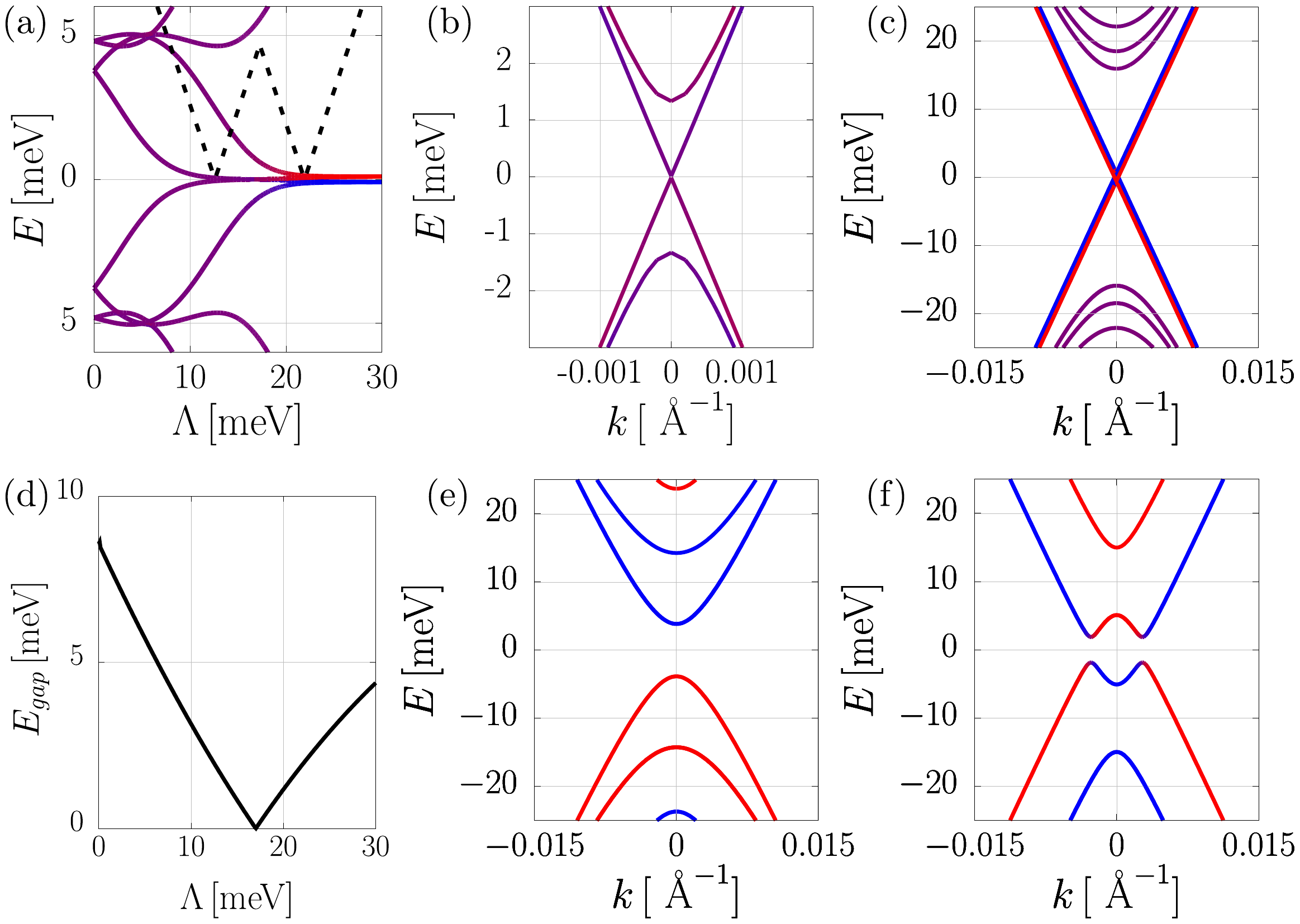}
    \caption{\label{fig:Energy-Bands}
    (a) Low-energy states at $k=0$ and (b)-(c) energy spectrum for an infinitely long thin film with $\mu=0$. The black dashed line in (a) stands for the $k=0$ gap, while the band structures are computed with (b) $\Lambda=15$ meV and (c) $\Lambda=30$ meV.
     (d) Energy gap at $k=0$ and (e)-(f) band structures for an infinite wire with $\mu=10$ meV.
    The band structures are obtained with (e) $\Lambda=10$ meV and (f) $\Lambda=30$ meV.
    Red and blue colours represent electron and hole modes, respectively, and purple denotes a superposition.
    }
\end{figure}

When confined along the $z$ direction, the MTI Hamiltonian~\eqref{eq:MTI_Hamiltonian} can be used to describe the physical properties of a 2D (thin film) or 1D (wire) geometry \cite{QAHE_in_MTIs, Colloquium_QAHE, Chiral_TSC_Half-Integer_Plateau, Quasi-1D-QAH-Majorana, QAH_Majorana_Platform}.
The Hamiltonian for a 2D system with particle-hole symmetry belongs to the D symmetry class; therefore, an integer invariant $\mathcal{N}$ characterizes the topological state of the two-dimensional slab \cite{Topological_Table, Topological_classification_symmetries}.
A chiral TSC with odd Chern invariant and unpaired Majorana modes can be realized in a 2D thin film starting from the quantum anomalous Hall (QAH) phase, which is routinely achieved in MTIs \cite{QAHE_in_MTIs, Observation_QAH_MTI, QAH_intrinsic_MTI}.
For $\mu=0$, the proximity pairing induces a novel region between the $\mathcal{N}=0$ trivial superconductor and the $\mathcal{N}=2$  QAH state \cite{Majorana_backscattering}.
In this intermediate region, the MTI thin film realizes an $\mathcal{N}=1$ TSC with \emph{unpaired} chiral Majorana modes on the edges \cite{Chiral_TSC, Chiral_TSC_Half-Integer_Plateau}.
The occurrence of this chiral TSC region is shown in Fig.~\ref{fig:Energy-Bands}(a), which displays the $k=0$ low-energy eigenvalues of Eq.~\eqref{eq:BdG_Hamiltonian} solved in the thin film geometry as a function of the Zeeman field $\Lambda$.
The black dashed line represents the $k=0$ bulk energy gap, showing the existence of two distinct transition points where topological phase transitions occur with the emergence of gapless edge modes within the bulk gap.
Figs.~\ref{fig:Energy-Bands}(b)-(c) display the full band structure of these phases: the first shows a single crossing of unpaired MCPSs which characterizes the $\mathcal{N}=1$ chiral topological superconductor, the second corresponds to the BdG quasiparticle spectrum of an $\mathcal{N}=2$ proximitized QAH system.

While a large thin film can realize different 2D topological superconducting states, a narrow MTI wire with $\mu \neq 0$ can be used to achieve a quasi-one-dimensional TSC with end-localized MBSs.
Since the effective BdG Hamiltonian of a QAH/SC heterostructure in a 1D geometry fits in the BDI symmetry class \cite{Quasi-1D-QAH-Majorana}, the topological properties of the system are characterized by an integer invariant \cite{Topological_Table, Topological_classification_symmetries} which discriminates between trivial $N_{BDI}=0$ and topological $N_{BDI}=1$ states with unpaired Majorana edge modes in finite-length systems.
In principle, even higher topological states can be realized, with $N_{BDI} \ge 2$ MBSs at the same end of the ribbon, protected by chiral symmetry.
However, the latter is broken in presence of disorder, and a pair of Majorana modes localized at the same end of the wire will fuse into a trivial fermion \cite{Kitaev_chain}.
Therefore, only the $(-1)^{N_{BDI}} = -1$ phases are topologically nontrivial in realistic samples, as a single \emph{unpaired} Majorana mode can be protected by particle-hole symmetry alone \cite{Quasi-1D-QAH-Majorana}.

The spectral gap at $k=0$ for an infinitely long wire with $\mu=10$ meV is shown in Fig.~\ref{fig:Energy-Bands}(d), where the closing and reopening of the energy gap signals a phase transition between trivial and topological states. The full band structures of the two distinct phases are depicted in Figs.~\ref{fig:Energy-Bands}(e)-(f).
It can be noted in Fig.~\ref{fig:Energy-Bands}(f) that the normal order of the energy bands around $k=0$ is inverted, indicating a nontrivial topology of the bulk and, as a consequence of the bulk-boundary correspondence, the presence of topologically protected MBSs at
the ends of wires with finite length \cite{Colloquium_TIs}.

\section{Antisymmetric Conductance}
\label{sec:Conductance}

\begin{table}
\begin{center}
\begin {tabular}{c|c|c|c}
\hline
\hline
S-Phase     & $G_1$ & $G_2$ & $G_t$  \\
\hline
$\mathcal{N}=0$     & 0                 & 0                     & 0 \\
$\mathcal{N}=1$     & $\alpha e^2/h$    & $(\alpha-1) e^2/h$    & $(2\alpha-1) e^2/h$  \\
$\mathcal{N}=2$     & $e^2/h$           & $-e^2/h$              & 0 \\
$N_{BDI}=0$         & 0                 & 0                     & 0 \\
$N_{BDI}=1$         & $2\alpha e^2/h$   & $2(\alpha-1) e^2/h$   & $2(2\alpha-1) e^2/h$ \\
ABS                 & $2\alpha e^2/h$   & 0                     & $2\alpha e^2/h$ \\
\hline
\hline
\end{tabular}
\end{center}
\caption{\label{tab:conductances}
Low-bias conductances $G_1, G_2$ and $G_t=G_1+G_2$ computed through Eqs. \eqref{eq:G_1}-\eqref{eq:G_2}.
The first column summarizes all the possible phases in the central S lead of the junction. Here, we considered a trivial ABS with perfect Andreev reflection on the left side of the junction.
The conductances are given for $\beta=1-\alpha$.
}
\end{table}

\begin{figure*}
    \centering
    \includegraphics[width=0.75\linewidth]{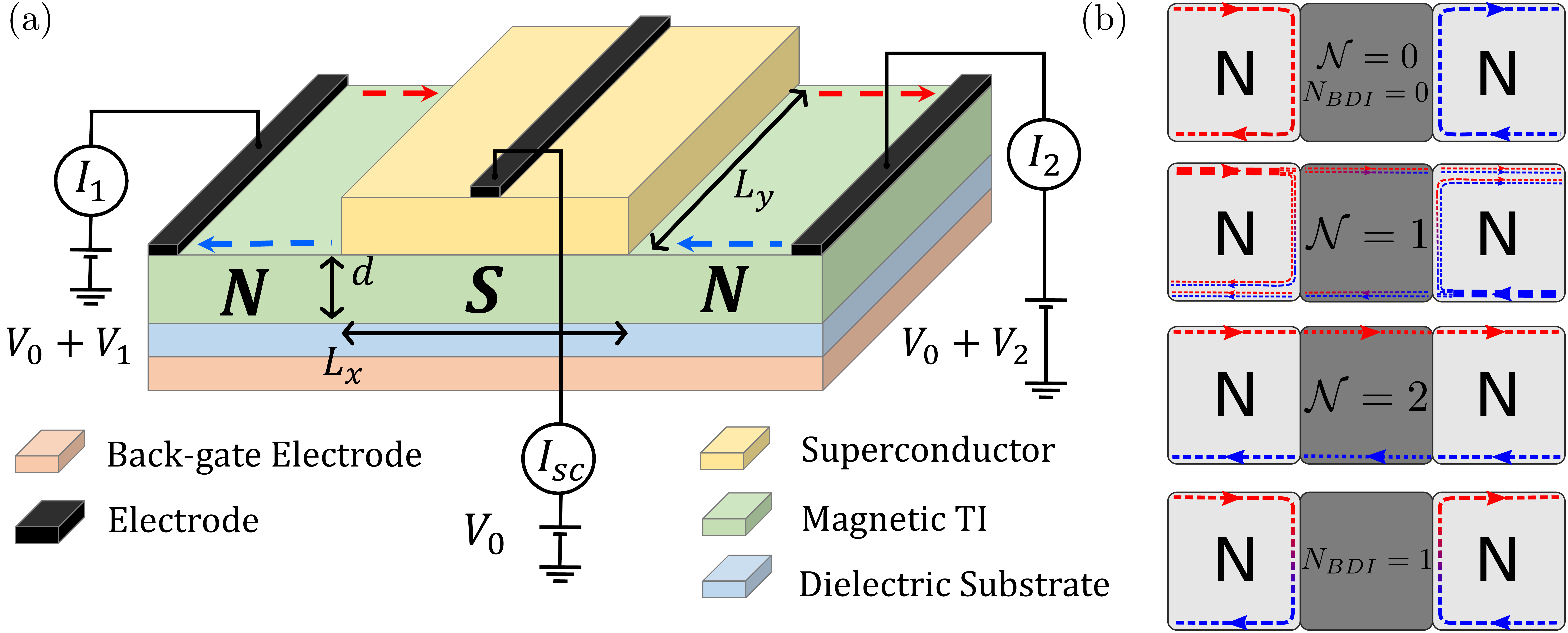}
    \caption{\label{fig:Setup}
    (a) Experimental setup proposed for the detection of topologically-protected Majorana modes. The potential $V_0=-e \mu$ is set by the back-gade electrode.
    (b) Sketches of the transmission processes at the interfaces of the junction for different superconducting phases in the central sector. Red and blue colors stand for electron and hole currents.
    }
\end{figure*}

Next, we analyse the electric transport through an NSN junction consisting of an MTI slab with central proximitized sector, exploring the regime in which the bias voltage drops asymmetrically over the left and right leads.
The experimental setup is schematically shown in Fig.~\ref{fig:Setup}(a).
The electric current $I_i$ in the normal terminals $i=1,2$ of a double junction can be computed as \cite{BTK, Lambert, Transport_Majorana_Nanowires}
\begin{equation}
    I_i = \int_0^{+\infty} dE \sum_a s_a\, \lbrack J_i^a (E) - K_i^a (E) \rbrack \,,
    \label{eq6}
\end{equation}
where $a \in \{e,h\}$ denotes electron and hole degrees of freedom, $s_{e,h}=\pm 1$, and
\begin{eqnarray}
\label{eq7}
    J_i^a (E) & = & \frac{e}{h} N_i^a(E)\, f_i^a(E) \,, \\
\label{eq8}
    K_i^a (E) & = & \frac{e}{h} \sum_{jb} P_{ij}^{ab}(E)\, f_j^b(E) \,,
\end{eqnarray}
are the incoming and outgoing fluxes of quasiparticles, respectively. The electric current is expressed in terms of the number of propagating modes in each terminal $N_i^a$ and the Fermi distribution functions $f_i^a$. Moreover, $P_{ij}^{ab}$ denotes the transmission probability of a quasiparticle of type $b \in \{e,h\}$ in lead $j$ to a quasiparticle of type $a$ in lead $i$, such that both normal ($a=b$) and Andreev ($a \neq b$) reflection ($i=j$) and transmission ($i \neq j$) are taken into account.
We define the differential conductance in the normal terminals of the double junction as
\begin{equation}
    G_{i} = \frac{\partial I_i}{\partial V} \,,
\end{equation}
where $V=V_1-V_2$ is the \emph{total} bias across the junction and $V_i$ is the voltage difference between the $i$-th lead and the central sector.
Here, we assume an asymmetric bias $V_1=\alpha V$ and $V_2=-\beta V$ with $0 \leq \alpha \leq 1$ and $\alpha+\beta=1$, such that the total bias between left and right terminals is fixed.
With this assumption, we can derive the following expressions for the conductance in the normal leads
\begin{eqnarray}
    G_1(V) & = & \alpha \frac{e^2}{h} \left\lbrack N_1^e(\alpha V) - P_{11}^{ee}(\alpha V) + P_{11}^{he}(\alpha V) \right\rbrack + \nonumber \\
    & + & \beta \frac{e^2}{h} \left\lbrack P_{12}^{hh}(\beta V) - P_{12}^{eh}(\beta V) \right\rbrack \,,
    \label{eq:G_1}\\
    G_2(V) & = & \beta \frac{e^2}{h}  \left\lbrack  -N_2^h(\beta V) - P_{22}^{eh}(\beta V) + P_{22}^{hh}(\beta V) \right\rbrack + \nonumber \\
    & + & \alpha \frac{e^2}{h} \left\lbrack P_{21}^{he}(\alpha V) - P_{21}^{ee}(\alpha V) \right\rbrack \,.
\label{eq:G_2}
\end{eqnarray}

The different possible transport processes in the junction are sketched in Fig.~\ref{fig:Setup}(b) and depend on the topological phase of the proximitized sector.
In the thin-film configuration, the scattering amplitudes can be merely inferred from the connection between edge modes in different sectors of the junction.
If the normal leads are held into a QAH state, a pair of zero-energy chiral modes run along the edges of the system, due to the particle-hole degeneracy of the $\Delta=0$ BdG Hamiltonian.
When $\mathcal{N}=2$, the superconducting sector is topologically equivalent to the QAH insulator in the normal terminals \cite{Chiral_TSC}. The chiral states run uninterruptedly through normal and proximitized leads, and the edge current is perfectly transmitted \cite{Majorana_backscattering, Andreev-oscillation-edge}.
Conversely, the $\mathcal{N}=0$ trivial superconductor does not support edge modes within the gap, and the boundary between the QAH phase and the proximitized region requires the occurrence of a gapless chiral state along the interface. This mode is responsible for the complete backscattering of the edge current flowing toward the superconductor \cite{Majorana_backscattering, Andreev-oscillation-edge}.
Finally, when $\mathcal{N}=1$, the proximitized sector supports a single unpaired Majorana mode on each edge. The injected modes from the QAH regions separate into two MCPSs at the interfaces of the junction, one is perfectly transmitted while the other is totally reflected.
Within the Blonder-Tinkham-Klapwijk formalism for the electric conductance in NSN junctions \cite{BTK, Lambert}, this process corresponds to equal probability of normal reflection, Andreev reflection, normal transmission and Andreev transmission \cite{Chiral_TSC_Half-Integer_Plateau, Majorana_backscattering}.

A similar framework can also be obtained for the wire geometry. 
When the width of the slab is smaller than the localization length of the edge modes, the QAH edge states are coupled into a single conducting channel, resembling a spinless metallic phase.
Since the proximitized sector realizes an effective 1D $p$-wave superconductor \cite{Quasi-1D-QAH-Majorana}, the interfaces between normal and superconducting MTI reproduce the physics of a NS junction between a normal metals and a $p$-wave superconductor \cite{Conductance_NS_Wire}.
In the $N_{BDI}=1$ topological state with end-localized MBSs, perfect Andreev reflection occurs for a bias lower than the energy gap \cite{Majorana_Andreev_Reflection}.
Conversely, in the $N_{BDI}=0$ trivial phase, the electric conductance is expected to vanish in the low-bias limit \cite{Conductance_NS_Wire}, meaning that the scattering processes are dominated by normal reflection.
Choosing appropriate values for the transmission probabilities $P_{ij}^{ab}$ in order to recover the scenarios above, the conductance on the two terminals of the junction can be easily computed from Eqs.~\eqref{eq:G_1}-\eqref{eq:G_2}.
Their values are summarized in Table~\ref{tab:conductances} for the different phases in the proximitized MTI and for a trivial ABS perfectly coupled to the left side of the junction in a wire geometry.
It can be noted that the total conductance $G_t =  G_1 + G_2 \neq 0$ only in presence of Majorana modes or ABSs, meaning that the currents in the N leads are different, being proportional to the fraction $\alpha, \beta$ of the total bias applied on the two sides of the junction.

We claim that the analysis of the total conductance $G_t$ as a function of bias splitting $\alpha$ can provide a useful criterion to rule out transport signatures from trivial Andreev processes. While not being a conclusive proof, an antisymmetric $G_t(\alpha)$ around $\alpha = 0.5$ would point towards Majorana modes because trivial Andreev levels are typically not constrained to such an antisymmetric profile.
The gapped superconductor and the proximitized QAH state exhibit a constant $G_t=0$, but, more generally, the emergence of trivial Andreev levels allow other $G_t(\alpha)$ trends, depending on how the ABS couples with the two interfaces of the junction.
For instance, in the aforementioned case of a trivial ABS perfectly coupled to the left lead only, the total conductance is antisymmetric around $\alpha=0$ (completely unbalanced bias splitting).
Quite remarkably, topological states with Majorana modes require $G_t(\alpha)$ to be \emph{antisymmetric} around $\alpha=0.5$ (equal bias splitting), because identical scattering amplitudes are expected at the two interfaces of the junction.
In spite of the same symmetry, a different ratio $G_t/G_0$, where $G_0 = e^2/h$ is the conductance quantum, characterizes different types of Majorana modes in the superconductor: in the case of a MCPS, $G_t/G_0 = (2 \alpha -1)$ due to all scattering probabilities
in Eqs.\ \eqref{eq:G_1}-\eqref{eq:G_2} being 0.25, while in presence of a MBS, $G_t/G_0 = 2(2 \alpha -1)$ indicates perfect Andreev reflection at the extremities of the proximitized sector.
Similar signals could be, in principle, obtained due to trivial ABSs, but the antisymmetry around $\alpha=0.5$ would in this case require fine-tuned equal conditions on the two interfaces of the junction.

The total conductance $G_t \neq 0$ is related to the existence of an electric current going to ground from the superconductor, which ensures charge conservation when $G_1 \neq G_2$ and the current injected on the left lead is different from the one flowing out on the right one.
This current can be easily detected through electric measurements, providing a measure of the electric conductance in the two normal leads, while maintaining the correlation between transport on the two interfaces of the junction.
We point out that, within our simplified model, the only current flowing through the $s$-wave superconductor is due to Cooper pairs originating in the proximitized MTI.
Indeed, for bias lower than the bulk gap, no quasiparticle modes can be excited, preventing unintended transmissions between the terminals of the junction.
We also neglected scattering processes occurring between the normal leads and the $s$-wave superconductor, since the presence of a physical interface between the two distinct materials would make suppress them compared to the scattering events which take place within the MTI slab.

\subsection*{Conductance Matrix}
The antisymmetric relation involving the total conductance $G_t$ can be expressed in the equivalent language of the conductance matrix, where the current-voltage relation reads
\begin{equation}
    \begin{pmatrix}
        I_1 \\
        I_2
    \end{pmatrix}
    =
    \begin{pmatrix}
        g_{11} & g_{12} \\
        g_{21} & g_{22}
    \end{pmatrix}
    \begin{pmatrix}
        V_1 \\
        V_2
    \end{pmatrix} \,.
\end{equation}
By considering 2-terminal transport between the N leads of our 3-terminal device (see Fig.~\ref{fig:Setup}), we can extract information about the Andreev processes taking place at the proximitized section of the MTI slab \cite{Conductance-Symmetries-2, Conductance-Symmetries-3}.
The conductance matrix elements are defined as $g_{ij} = \partial I_i / \partial V_j$ and can be distinguished into local ($i=j$) and nonlocal ($i \neq j$) components.
The conductance $G_{i}$ for the current in the $i=1,2$ terminal takes the form
\begin{equation}
\begin{split}
    G_1 & = \alpha\, g_{11} + (\alpha-1)\, g_{12} \,, \\
    G_2 & = \alpha\, g_{21} + (\alpha-1)\, g_{22} \,,
\end{split}
\end{equation}
and the total conductance can be written as
\begin{equation}
    G_t = -( g_{12} + g_{22} ) + \alpha (g_{11}+g_{12}+g_{21}+g_{22}) \,.
\end{equation}
Therefore, in terms of local and nonlocal matrix elements the antisymmetric condition around $\alpha=0.5$ can be written explicitly as
\begin{equation}\label{eq:symmetry-condition}
    g_{11} - g_{12} = g_{22} - g_{21} \,,
\end{equation}
meaning that the difference between local and nonlocal conductances must be the same
in both terminals.

The novelty of our approach compared with previous works is that we focus on the symmetry of $G_t$ around the limit $\alpha=0.5$ of equally-split bias voltage.
For instance, in the case of a superconducting wire with end-localized MBSs, the local conductances $g_{11}$ and $g_{22}$ are usually measured separately, looking for correlations of $2e^2/h$ zero-bias peaks on the two sides of the nanowire.
In our framework of asymmetrically split bias, these measurements are equivalent to $\alpha=1$ and $\alpha=0$, respectively.
Our work shows that measuring $G_t$ continuously as a function of $\alpha$ can provide a more robust transport signature.
Similarly, in a proximitized MTI thin film, a single transport measurement of $G_t$ in a junction with equal bias split $\alpha=0.5$ has been already proved to be incapable of detecting Majorana chiral propagating states \cite{Retracted_MCPS, Absence_Evidence}.
Determining $G_t$ with different bias configurations can provide more information on the phase of the proximitized sector.

\section{Numerical Results}
\label{sec:Numerical}

\begin{figure}
    \centering
    \includegraphics[width=\linewidth]{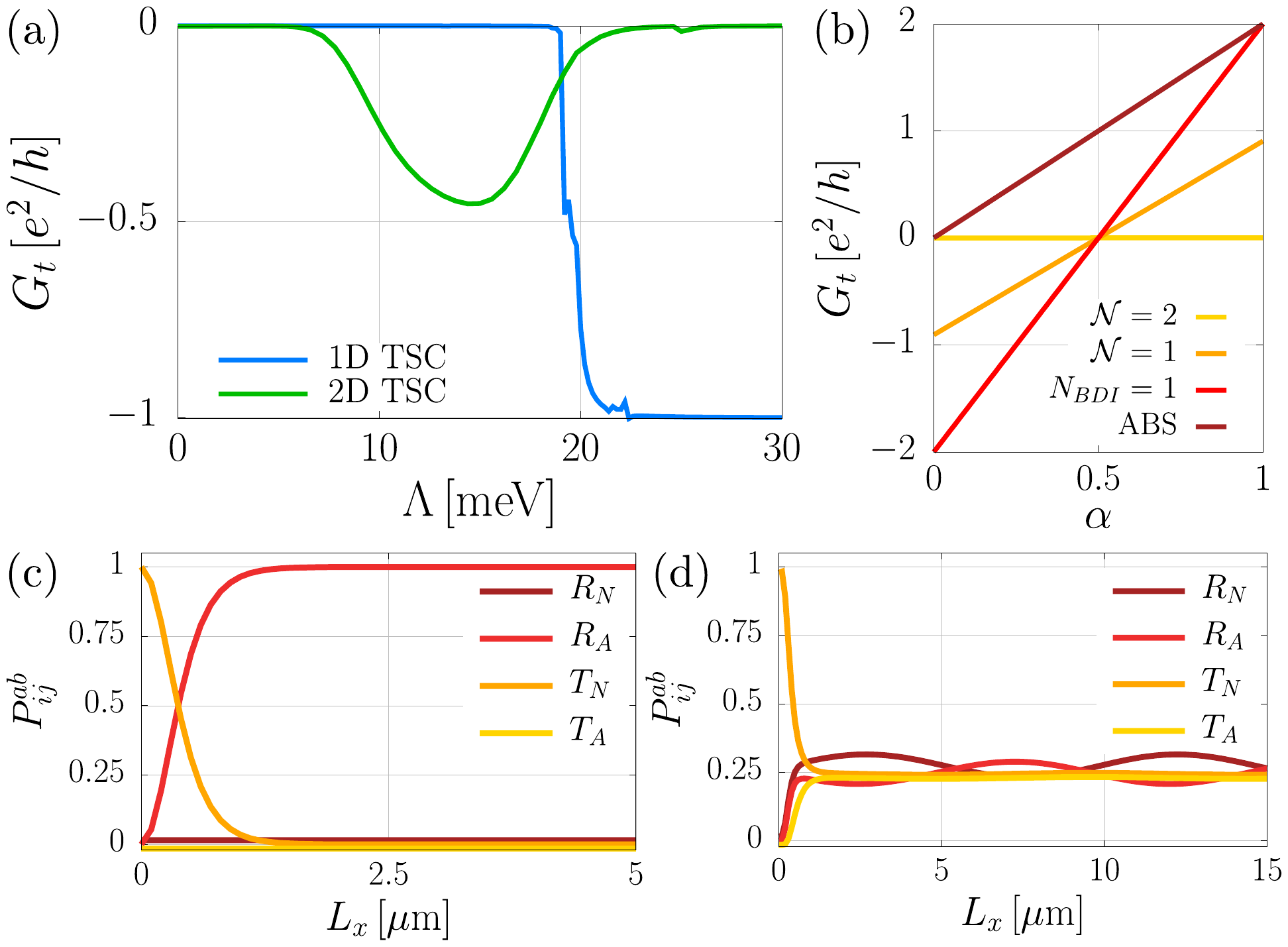}
    \caption{\label{fig:Conductances}
    Conductance $G_t$ computed in the NSN junction as a function of (a) magnetization and (b) bias split. In the left panel $\alpha=0.25$ and the blue (green) line stands for the wire (thin film) geometry.
    The ABS is modelled adding a barrier on the right side of the junction with the proximitized sector in the $N_{BDI}=1$ phase.
    (c)-(d) Transmission amplitudes $P_{ij}^{ab}$ for the left terminal of the junction as a function of the length $L_x$ of the central proximitized sector.
    The probabilities are computed for a (c) $N_{BDI}=1$ superconductor with MBSs and a (d) $\mathcal{N}=1$ superconductor with MCPSs.
    In all the pictures, the total bias $V=0.1$ meV is chosen within the bulk gap.
    The values of $\Lambda$ in panels (b),(c) and (d) are chosen according to Fig.~\ref{fig:Energy-Bands} to reproduce the different TSCs.
    }
\end{figure}

We simulated a NSN junction between proximitized and normal MTI slabs using a complex band structure approach. This numerical technique allows us to describe not only propagating modes with $k \in \mathbb{R}$, but also evanescent states (like end-localized MBSs) originating in the superconductor, which are related to complex longitudinal wavenumbers $k \in \mathbb{C}$.
Within this framework, the electron wavefunction in a homogeneous sector of the junction can be chosen as a superposition of transverse wavefunctions $\Psi_k (y,z; \eta)$ with a proper wavenumber $k$.
The full wavefunction takes the generic form
\begin{equation}
\Psi (x,y,z; \eta)  =
\sum_k{c_k} \, \Psi_k (y,z; \eta) \, e^{ik x}
\end{equation}
where $c_k$ are complex numbers and $\eta = \{\sigma, \tau, \delta \}$ represents the set of spin $\sigma$, orbital index $\tau$ and particle-hole type $\delta$.
Notably, the wavefunction for a finite-length sector of the junction can be constructed from the set of bulk wavenumbers and coefficients $\{k, c_k \}$ obtained in a full translational-invariant system.
Furthermore, while a grid discretization is required for the confined dimensions $y$ and $z$,  the dependence of $\Psi (x,y,z; \eta)$ along the longitudinal axis $x$ is parametric, and enables to describe a junction of any length $L_x$.
Further details about the technique are given in Sec.~\ref{sec:Method}.

We computed numerically the conductances $G_1, G_2$ and the sum $G_t=G_1+G_2$ in the NSN junction with a magnetic TI in the wire and thin film configurations, reproducing the properties of 1D and 2D topological superconductors, respectively.
Fig.~\ref{fig:Conductances}(a) displays $G_t$ versus the magnetization of the MTI for an asymmetric bias $\alpha=0.25$. In the thin film geometry, a region with $G_t \neq 0$ distinguishes the $\mathcal{N}=1$ chiral TSC from the $\mathcal{N}=0$ trivial superconductor and the  $\mathcal{N}=2$ QAH phase, where $G_t=0$ denotes that the electric currents in the two terminals are equal and opposite independently of the bias split.
For the chosen $\alpha$, the conductance for $\mathcal{N}=1$ was expected to be quantized at $G_t=-e^2/2h$, which is roughly the value reached in the nontrivial region with MCPSs.
Similarly, in the wire geometry a plateau  $G_t=-e^2/h$ characterizes the $N_{BDI}=1$ nontrivial phase, while the $N_{BDI}=0$ gapped superconductor exhibits $G_t=0$.
Fig.~\ref{fig:Conductances}(b) shows $G_t$ as a function of the split parameter $\alpha$
for all the  nontrivial phases realized by the proximitized MTI.
A trivial ABS is also simulated in the wire geometry as an $N_{BDI}=1$ superconductor with an insulating barrier on the right side of the junction.
Here, the values of the conductance are in perfect agreement with our prediction in Table~\ref{tab:conductances}.
We emphasize here that a symmetrically distributed bias $\alpha=0.5$ is insufficient to discriminate a TSC from the trivial state and the proximitized QAH phase, since this particular configuration implies always $G_t=0$.

The lower panels in Fig.~\ref{fig:Conductances} display the amplitudes $P_{ij}^{ab}$ for all the scattering processes occurring on the left interface of the junction, i.e., normal reflection $R_N$, Andreev reflection $R_A$, normal transmission $T_N$ and Andreev transmission $T_A$.
The figures correspond to a $N_{BDI}=1$ topological superconducting wire with unpaired MBSs in (c) and a $\mathcal{N}=1$ TSC thin film with MCPSs in (d).
The former shows that, when the junction is sufficiently large to prevent transmission by evanescent modes, the injected electron undergoes perfect Andreev reflection $R_A=1$ in presence of MBSs.
The latter indicates that due to MCPSs, normal and Andreev transmission and reflection occur with equal probability $R_N=R_A=T_N=T_A=0.25$.
Oscillations around the expected plateaus are due to the interference between back-scattered chiral modes from the two interfaces of the double junction, resulting in an interferometric behaviour \cite{Conductance_oscillations_MCPS}.
For all the above results, the total bias across the junction is $V=0.1$ meV, which is always lower than the bulk energy gap.
Such low bias ensures that no bulk modes are activated in the proximitized sector and that the injected electrons and holes interact in the condensate only with topologically-protected Majorana boundary states.
Indeed, the proposed framework does not hold for higher bias, which implies interaction with multiple active modes in the superconductor.

All the numerical simulations are obtained without taking into account the effect of disorder in the system. 
Even tough this may seem a crude approximation, in the 2D thin film the electric current is transported through the junction by topologically-protected chiral fermionic or Majorana edge modes.
These electronic states are known to be insensitive to weak disorder, because no energy modes are available for backscattering \cite{Colloquium_TIs}.
Conversely, in the regime of a narrow ribbon opposite edge states are strongly coupled and it is difficult to maintain the ballistic nature of the chiral channels.
Despite the fragility of the QAH edge states, the MBSs arising in the quasi-1D topological phase are expected to be robust against weak disorder, maintaining well-quantized zero-bias peaks in tunneling spectroscopy \cite{Robust_MBSs}.
Therefore, with asymmetric bias splitting, we expect the low-bias conductance to keep its antisymmetric behaviour in the presence of weak disorder.

\section{Conclusion}
\label{sec:Conclusion}

In summary, an asymmetric bias voltage drop applied across an NSN junction provides a useful, but not conclusive, criterion to rule out conductance signals produced by trivial Andreev levels in MTI slabs with a central proximitized section.
We showed that the antisymmetry of the conductance $G_t$ with respect to the point of equal bias splitting ($\alpha=0.5$) is a necessary condition for topologically-protected Majorana modes in normal-superconductor junctions.
Detailed model calculations for a narrow (wire-like) and a wide (film-like) slab, hosting MBSs and MCPSs respectively, are shown to support our conclusions.
Our results will be useful for the experimental detection of the elusive Majorana quasiparticles, contributing to the progress towards a solid platform for quantum computing.

\begin{acknowledgments}
This project is financially supported by the QuantERA grant MAGMA, by the National Research Fund Luxembourg under the grant INTER/QUANTERA21/16447820/MAGMA, by the German Research Foundation under grant 491798118, by MCIN/AEI/10.13039/501100011033 under project PCI2022-132927, and by the European Union NextGenerationEU/PRTR.
L.S.~acknowledges support from Grants No.~PID2020-117347GB-I00 funded by MCIN/AEI/10.13039/501100011033 and No.~PDR2020-12 funded by GOIB.
K.M.~acknowledges the financial support by the Bavarian Ministry of Economic Affairs, Regional Development and Energy within Bavaria’s High-Tech Agenda Project "Bausteine für das Quantencomputing auf Basis topologischer Materialien mit experimentellen und theoretischen Ansätzen" (grant allocation no.\ 07 02/686 58/1/21 1/22 2/23).
\end{acknowledgments}

\appendix

\section{Numerical Method}
\label{sec:Method}

Our numerical results are obtained using a grid discretization of the continuum
Hamiltonian~\eqref{eq:BdG_Hamiltonian}. States with real $k$, i.e., propagating modes like those in
Fig.~\ref{fig:Energy-Bands}, can be directly obtained by matrix diagonalization of the
corresponding BdG energy eigenvalue problem
\begin{equation}
\label{eqB1}
\mathcal{H}_{BdG}(k)\,\Psi = E\,\Psi\; .
\end{equation}
However, transport in non-translation-invariant systems like the NSN junction requires also more general (evanescent) states described by a complex wave number $k$.
We modelled this case adapting the complex-band-structure approach
discussed in Refs.~\cite{Osca19,Ben21}.

We first rewrite the Hamiltonian by explicitly separating the $k$-dependent terms
as $\mathcal{H}_{BdG}={\cal A}+{\cal B} k +{\cal C}k^2$.
Then, defining an enlarged wave function $(\Psi_1,\Psi_2)^T=(\Psi,k\Psi)^T$ it is possible to reformulate the energy eigenvalue problem in Eq.~\eqref{eqB1} into a $k$-eigenvalue problem. After some straightforward algebra,
this reads
\begin{equation}
\label{eqB2}
\left(
\begin{array}{cc}
0 & 1 \\
-{\cal C}^{-1}({\cal A}-E) &
-{\cal C}^{-1}{\cal B}
\end{array}
\right)
\left(
\begin{array}{c}
\Psi_1 \\
\Psi_2
\end{array}
\right)
=
k
\left(
\begin{array}{c}
\Psi_1 \\
\Psi_2
\end{array}
\right)\; .
\end{equation}
In terms of the original parameters of the MTI Hamiltonian Eq.~\eqref{eq:BdG_Hamiltonian}
it is
\begin{eqnarray}
{\cal A} &=& C_0 - C_\perp k_y^2 - C_z k^2_z \nonumber\\
&+& (M_0 - M_\perp k^2_y - M_z k^2_z ) \tau_z \nonumber\\
&+& ( A_\perp k_y \sigma_y + A_z k_z \sigma_z) \tau_x\, ,\\
{\cal B} &=& A_\perp \sigma_x \tau_x\, ,\\
{\cal C} &=& -(C_\perp + M_z \tau_z) \, .
\end{eqnarray}
Note that the eigenvalue problem~\eqref{eqB2} requires non-Hermitian matrix solvers to include the possibility of complex wave numbers $k$.

In the modelling of the NSN double junction, we first solve the matrix version of Eq.~\eqref{eqB2}
for a large set of modes $\{k^{(a)},\Psi^{(a)}_k\}$
in each sector, where $a=L,C,R$, refers to left, centre and right, respectively.
The wave function is then represented by a collection of input (output) amplitudes
$a_k^{(a)}$ $(b_k^{(a)})$ in each part as
\begin{equation}
\begin{split}
\Psi^{(a)} (x,y,z; \eta) & =
\sum_k{a_k^{(a)}} \, \Psi_k^{(a)} (y,z; \eta) \, e^{ik^{(a)}(x-x_k^{(a)})} \\
& + \sum_k{b_k^{(a)}} \, \Psi_k^{(a)} (y,z; \eta) \, e^{ik^{(a)}(x-x_k^{(a)})} \; ,
\label{eqB6}
\end{split}
\end{equation}
where $\eta = \{\sigma, \tau, \delta \}$ represent the set of spin $\sigma$, orbital index $\tau$ and particle-hole type $\delta$, and the input/output character of each mode $\Psi_k$ is determined according to the sign of its probability flux
\begin{equation}
    I_k=\langle\Psi_k | \partial H/\partial k_x|\Psi_k\rangle \,.
\end{equation}
The $e^{i k^{(a)} x_k^{(a)}}$ factors in Eq.~\eqref{eqB6} are a gauge choice that helps avoid numerical instabilities \cite{Ben21}.

Due to truncation, the total number of unknowns
$\{b_k^{(L)},b_k^{(C)},b_k^{(R)}\}$
is finite and their values must be fixed by imposing continuity of the wave function and its $x$-derivative at the two interfaces
$x=x_1$ and $x=x_2$. In practice, those equations are projected onto the total discrete set of complex modes by means of the overlap matrices
\begin{equation}
{\cal M}_{k'k}^{(ab)}
=
\sum_{\sigma \tau \delta}{}
\int{dydz\,
\Psi_{k'}^{a*}(y,z; \eta) \,
\Psi_k^{b}(y,z; \eta)
}\; .
\end{equation}

In detail, the linear system reads
\begin{widetext}
\begin{equation}
\begin{array}{rrll}
\displaystyle\sum_{k^{(L)}}{
{\cal M}_{k'k}^{(a L)}\,
{b_k^{(L)}}
}
&
-
\displaystyle\sum_{k^{(C)}}{
{\cal M}_{k'k}^{(a C)}\,
e^{i k^{(C)}(x_1-x_k^{(C)})}\,
{b_k^{(C)}}
}
&=
-\displaystyle\sum_{k^{(L)}}{
{\cal M}_{k'k}^{(a L)}\,
{a_k^{(L)}}
}\; ,
&
\;\; {\rm if}\;
\left\{
\begin{array}{c}
a=L\\
x_{k'}^{(a)}=x_1
\end{array}
\right.
\; , \\
\rule{0cm}{0.8cm}
\displaystyle\sum_{k^{(L)}}{
{\cal M}_{k'k}^{(a L)}\,
k^{(L)}
{b_k^{(L)}}
}
&
-
\displaystyle\sum_{k^{(C)}}{
{\cal M}_{k'k}^{(a C)}\,
e^{i k^{(C)}(x_1-x_k^{(C)})}\,
k^{(C)}
{b_k^{(C)}}
}
&=
-\displaystyle\sum_{k^{(L)}}{
{\cal M}_{k'k}^{(a L)}\,
k^{(L)}
{a_k^{(L)}}
}\; ,
&
\;\; {\rm if}\;
\left\{
\begin{array}{c}
a=C\\
x_{k'}^{(a)}=x_1
\end{array}
\right.
\; , \\
\rule{0cm}{0.8cm}
\displaystyle\sum_{k^{(R)}}{
{\cal M}_{k'k}^{(a R)}\,
k^{(R)}
{b_k^{(R)}}
}
&
-
\displaystyle\sum_{k^{(C)}}{
{\cal M}_{k'k}^{(a C)}\,
e^{i k^{(C)}(x_2-x_k^{(C)})}\,
k^{(C)}
{b_k^{(C)}}
}
&
=
-\displaystyle\sum_{k^{(R)}}{
{\cal M}_{k'k}^{(a R)}\,
k^{(R)}
{a_k^{(R)}}
}\; ,
&
\;\;{\rm if}\;
\left\{
\begin{array}{c}
a=C\\
x_{k'}^{(a)}=x_2
\end{array}
\right.
\; , \\
\rule{0cm}{0.8cm}
\displaystyle\sum_{k^{(R)}}{
{\cal M}_{k'k}^{(a R)}\,
{b_k^{(R)}}
}
&
-
\displaystyle\sum_{k^{(C)}}{
{\cal M}_{k'k}^{(a C)}\,
e^{i k^{(C)}(x_2-x_k^{(C)})}\,
{b_k^{(C)}}
}
&
=
-\displaystyle\sum_{k^{(R)}}{
{\cal M}_{k'k}^{(a R)}\,
{a_k^{(R)}}
}\; ,
&
\;\;{\rm if}\;
\left\{
\begin{array}{c}
a=R\\
x_{k'}^{(a)}=x_2
\end{array}
\right.\; .
\label{eq4n}
\end{array}
\end{equation}

By solving Eq.~\eqref{eq4n}
with $a_k^{(a)} = 1$ for a particular input propagating mode, with all other inputs vanishing, we obtain a particular input/output transmission probability $p_{k'k}^{a'a}=|b_{k'}^{a'}|^2$.
The sum of all these individual probabilities discriminating their electron/hole character in the normal leads finally yields the total probabilities defined in Sec.~\ref{sec:Conductance}
\begin{equation}
P_{ji}^{he}=\sum_{k_h k_e}{p_{k_h k_e}^{ji}} \,.
\end{equation}
A good control of the model truncations, regarding grid size and number of complex modes, is given by the flux conservation, which we typically require to be better than 1\%.

\section{Derivation of the Differential Conductances}

We derive here the equations for the conductances $G_1$ and $G_2$ given in Eqs.~\eqref{eq:G_1} and \eqref{eq:G_2}.
In our approach \cite{Lambert}, the bias dependence is contained in Fermi energy distribution functions for electrons and holes  injected from far-distant reservoirs into the N leads
\begin{equation}
    f_i^a(E) =
    \begin{cases}
    \frac{1}{1+e^{\left( E - e V_i \right)/k_B T}}  & \text{if } a=e\;, \\
    \frac{1}{1+e^{\left( E + e V_i \right)/k_B T}}  & \text{if } a=h\; ,
    \end{cases}
\end{equation}
with $V_i$ being the voltage difference between the $i$-th reservoir and the MTI slab.
By making explicit the sum over the quasiparticle types of
Eq.~\eqref{eq6} and using Eqs.~\eqref{eq7} and \eqref{eq8}, the electric current can be rewritten as
\begin{equation}
\begin{split}
    I_i & = \frac{e}{h} \int_0^{+\infty} dE \left \lbrack J_i^e - K_i^e - J_i^h + K_i^h \right \rbrack
    = \frac{e}{h} \int_0^{+\infty} dE \left \lbrack
    N_i^e f_i^e -\sum_{jb} P_{ij}^{eb} f_j^b - N_i^h f_i^h + \sum_{jb} P_{ij}^{hb} f_j^b
    \right \rbrack \\
    & = \frac{e}{h} \int_0^{+\infty} dE\,  \left \lbrack
    N_i^e f_i^e -\sum_j \left( P_{ij}^{ee} f_j^e + P_{ij}^{eh} f_j^h  \right)
    - N_i^h f_i^h + \sum_j \left( P_{ij}^{he} f_j^e + P_{ij}^{hh} f_j^h \right)
    \right \rbrack \,,
\end{split}
\end{equation}
where for simplicity we omitted the energy dependence.
Expanding the sum over the terminals $j=1,2$ we can write the electric current into the two leads of the junction as
\begin{align}
    I_1 & = \frac{e}{h} \int_0^{+\infty} dE \Biggl\{
    \left\lbrack  N_1^e - P_{11}^{ee} + P_{11}^{he}  \right\rbrack f_1^e 
     + \left\lbrack -N_1^h - P_{11}^{eh} + P_{11}^{hh}  \right\rbrack f_1^h
     + \left\lbrack P_{12}^{he} - P_{12}^{ee} \right\rbrack  f_2^e +
    \left\lbrack P_{12}^{hh} - P_{12}^{eh} \right\rbrack  f_2^h
    \Biggr\} \,, \notag \\
    I_2 & = \frac{e}{h} \int_0^{+\infty} dE \Biggl\{
    \left\lbrack  N_2^e - P_{22}^{ee} + P_{22}^{he}  \right\rbrack f_2^e 
    + \left\lbrack -N_2^h - P_{22}^{eh} + P_{22}^{hh}  \right\rbrack f_2^h 
    + \left\lbrack P_{21}^{he} - P_{21}^{ee} \right\rbrack  f_1^e +
    \left\lbrack P_{21}^{hh} - P_{21}^{eh} \right\rbrack  f_1^h
    \Biggr\} \,.
\end{align}
We assume that the bias is asymmetrically distributed as $V_1=\alpha V$ and $V_2=-\beta V$ with $0 \leq \alpha \leq 1$ and $\beta = 1-\alpha$ such that the total voltage drop across the NSN junction is fixed to $V_1-V_2 = V$. In the zero-temperature limit the Fermi functions take the form of step functions
\begin{align}
    f_1^{e,h} & = \frac{1}{1+e^{\left( E \mp e \alpha V \right)/k_B T}} \xrightarrow[T \to 0]{} \Theta( E \mp \alpha e V )
\end{align}
for the left terminal of the junction and analogously for the right junction.
The expressions of the currents in the two terminals can thus be simplified as
\begin{align}
    I_1 \label{eq:I_1}
    & = \frac{e}{h} \int_0^{\alpha e V} dE  \left\lbrack  N_1^e - P_{11}^{ee} + P_{11}^{he}  \right\rbrack +
    \frac{e}{h} \int_0^{\beta e V} dE \left\lbrack P_{12}^{hh} - P_{12}^{eh} \right\rbrack \,, \\
    I_2 \label{eq:I_2}
    & = \frac{e}{h} \int_0^{\beta e V} dE  \left\lbrack  -N_2^h - P_{22}^{eh} + P_{22}^{hh} \right\rbrack +
    \frac{e}{h} \int_0^{\alpha e V} dE \left\lbrack P_{21}^{he} - P_{21}^{ee} \right\rbrack \,,
\end{align}
and the differential conductance can be computed as the derivative of Eqs.~\eqref{eq:I_1}-\eqref{eq:I_2} with respect to the total bias $V$ across the junction, leading to
\begin{align}
    \label{app-eq:G_1} G_1(V) = \frac{\partial I_1}{\partial V} & = \alpha \frac{e^2}{h} \left\lbrack N_1^e(\alpha V) - P_{11}^{ee}(\alpha V) + P_{11}^{he}(\alpha V) \right\rbrack 
    + \beta \frac{e^2}{h} \left\lbrack P_{12}^{hh}(\beta V) - P_{12}^{eh}(\beta V) \right\rbrack \,, \\
    \label{app-eq:G_2} G_2(V) = \frac{\partial I_2}{\partial V} & = \beta \frac{e^2}{h}  \left\lbrack  -N_2^h(\beta V) - P_{22}^{eh}(\beta V) + P_{22}^{hh}(\beta V) \right\rbrack 
    + \alpha \frac{e^2}{h} \left\lbrack P_{21}^{he}(\alpha V) - P_{21}^{ee}(\alpha V) \right\rbrack \,.
\end{align}
\end{widetext}
The left-terminal conductance~\eqref{app-eq:G_1} is given by the number of injected electrons $N_1^e$, the normal $P_{11}^{ee}$ and Andreev $P_{11}^{he}$ reflection amplitudes for electrons injected in lead 1, and the normal $P_{12}^{hh}$ and Andreev $P_{12}^{eh}$ transmission amplitudes for holes injected in lead 2.
Similarly, the right-terminal conductance Eq. \eqref{app-eq:G_2} is given by the number of injected holes $N_2^h$, normal $P_{22}^{hh}$ and Andreev $P_{22}^{eh}$ reflection amplitudes for holes injected in lead 2, and normal $P_{21}^{ee}$ and Andreev $P_{21}^{he}$ transmission amplitudes for electrons injected in lead 1.
The number of injected quasiparticles $N_i^a$ and the values of the transmission amplitudes $P_{ij}^{ab}$
in the low-bias scenario described in the main article are given in Tables~\ref{tab:G_1}-\ref{tab:G_2} for all the topological phases of the superconducting sector.
Table~\ref{app-tab:conductances} summarizes the corresponding values of the conductances $G_1,G_2$ and their sum $G_t=G_1+G_2$.

\begin{table}
\begin{center}
\begin {tabular}{l|c|c|c|c|c||c}
\hline
\hline
S-Phase     & $N_1^e$ & $P_{11}^{ee}$ & $P_{11}^{he}$ & $P_{12}^{hh}$ & $P_{12}^{eh}$ & $G_1$ \\
\hline
$N_{BDI}=0$             & 1     & 1     & 0    & 0    & 0     & 0               \\
$N_{BDI}=1$ (MBS)       & 1     & 0     & 1    & 0    & 0     & $2\alpha e^2/h$ \\
$\mathcal{N}=0$         & 1     & 1     & 0    & 0    & 0     & 0               \\
$\mathcal{N}=1$ (MCPS)  & 1     & 0.25  & 0.25 & 0.25 & 0.25  & $\alpha e^2/h$  \\
$\mathcal{N}=2$ (QAH)   & 1     & 0     & 0    & 1    & 0     & $e^2/h$ \\
\hline
\hline
\end{tabular}
\end{center}
\caption{\label{tab:G_1}
Transmission amplitudes and number of electronic modes required to compute the conductance $G_1$ through Eq. \eqref{app-eq:G_1}. The values are given for all the possible topological phases which can be found in the central superconducting sector of the NSN junction.
The conductances are given taking into account that $\beta=\alpha-1 $.
}
\end{table}

\begin{table}
\begin{center}
\begin {tabular}{l|c|c|c|c|c||c}
\hline
\hline
S-Phase     & $N_2^h$ & $P_{22}^{eh}$ & $P_{22}^{hh}$ & $P_{21}^{he}$ & $P_{21}^{ee}$ & $G_2$ \\
\hline
$N_{BDI}=0$             & 1     & 0     & 1    & 0    & 0     & 0                   \\
$N_{BDI}=1$ (MBS)       & 1     & 1     & 0    & 0    & 0     & $2(\alpha-1) e^2/h$ \\
$\mathcal{N}=0$         & 1     & 0     & 1    & 0    & 0     & 0                   \\
$\mathcal{N}=1$ (MCPS)  & 1     & 0.25  & 0.25 & 0.25 & 0.25  & $(\alpha-1) e^2/h$  \\
$\mathcal{N}=2$ (QAH)   & 1     & 0     & 0    & 0    & 1     & $-e^2/h$            \\
\hline
\hline
\end{tabular}
\end{center}
\caption{\label{tab:G_2}
Transmission amplitudes and number of hole modes required to compute the conductance $G_2$ through Eq. \eqref{app-eq:G_2}. The values are given for all the possible topological phases which can be found in the central superconducting sector of the NSN junction. The conductances are given taking into account that $\beta=\alpha-1 $.
}
\end{table}

\begin{table}
\begin{center}
\begin {tabular}{l|c|c|c}
\hline
\hline
S-Phase     & $G_1$ & $G_2$ & $G_t$  \\
\hline
$N_{BDI}=0$             & 0                 & 0                  & 0 \\
$N_{BDI}=1$ (MBS)       & $2\alpha e^2/h$   & $2(\alpha-1) e^2/h$  & $2(2\alpha-1) e^2/h$ \\
$\mathcal{N}=0$         & 0                 & 0                  & 0 \\
$\mathcal{N}=1$ (MCPS)  & $\alpha e^2/h$    & $(\alpha-1) e^2/h$   & $(2\alpha-1) e^2/h$  \\
$\mathcal{N}=2$ (QAH)   & $e^2/h$           & $-e^2/h$           & 0 \\
\hline
\hline
\end{tabular}
\end{center}
\caption{\label{app-tab:conductances}
Conductances $G_1, G_2$ and their sum $G_t=G_1+G_2$ computed through Eqs.~\eqref{app-eq:G_1}-\eqref{app-eq:G_2} using the transmission probabilities given in Tables~\ref{tab:G_1}-\ref{tab:G_2}. The sum of the conductance on the two terminal is non-zero only in presence of topologically-protected Majorana modes. Furthermore, the value of $G_t$ discriminates between end-localized MBSs and dispersive MCPSs.
}
\end{table}

\section{Role of an Interface Barrier}

\begin{table*}
\begin{center}
\begin {tabular}{l|c|c|c|c|c||c|c|c}
\hline
\hline
S-Phase     & $N_1^e$ & $P_{11}^{ee}$ & $P_{11}^{he}$ & $P_{12}^{hh}$ & $P_{12}^{eh}$ & $G_1$ & $G_2$ & $G_t$ \\
\hline
$N_{BDI}=1$ (MBS)       & 1     & 0     & 1    & 0    & 0     & $2\alpha e^2/h$ & 0 & $2\alpha e^2/h$ \\
$\mathcal{N}=1$ (MCPS)  & 1     & 0.5   & 0.5  & 0    & 0     & $\alpha e^2/h$  & 0 & $\alpha e^2/h$ \\
$\mathcal{N}=2$ (QAH)   & 1     & 1     & 0    & 0    & 0     & 0               & 0 & 0 \\
\hline
\hline
\end{tabular}
\end{center}
\caption{\label{tab:G_barrier}
Transmission amplitudes for the scattering processes on the left interface and conductances $G_1,G_2$ and $G_t$ assuming an insulating barrier on the right side of the system.
The values are given for all the topologically nontrivial cases.
}
\end{table*}

We consider in this section the role of an interface barrier between the central proximitized (S) sector  and the right (N) lead. That is, an NSN'N  structure where N' represents a slab of a normal MTI material without any propagating modes. The presence of N' breaks the left-right symmetry
with respect to the central sector and, depending on the
barrier transparency, it will affect the electric connection to the right side. A small barrier length
mimics interface disorder, while a large barrier length corresponds to the complete electric insulation.

We show here that the presence of a barrier does not change our conclusions about the relevance of $G_t$.
Indeed, despite the fact that the value of the conductance $G_1$ and $G_2$ may change, the total conductance keeps its meaning, with $G_t \neq 0$ as long as Andreev processes take place in the junction.
The different cases can be clearly understood when a completely insulating barrier is introduced, for instance, on the right side of the system: as the right lead is electrically disconnected from the proximitized MTI, $G_2=0$ regardless of the topological phase realized in the proximitized sector.
In an $N_{BDI}=1$ superconducting wire, perfect Andreev reflections occurs on the left interface of the junction due to interaction with MBS.
In an $\mathcal{N}=1$ TSC film, the electron is completely reflected, since the transmission to the right side is prevented by the barrier. Normal and Andreev processes take place with same probability.
In an analogous way, in an $\mathcal{N}=2$ TSC, the electrons are perfectly reflected from the barrier, and no Andreev processes take place in the junction.
The conductances $G_1$ and $G_2$ can be easily computed through Eqs. \eqref{eq:G_1}-\eqref{eq:G_2}. Their values, together with the transmission amplitudes for the left interface of the junction, are summarized in Table~\ref{tab:G_barrier}.

\begin{figure}
    \centering
    \includegraphics[width=\linewidth]{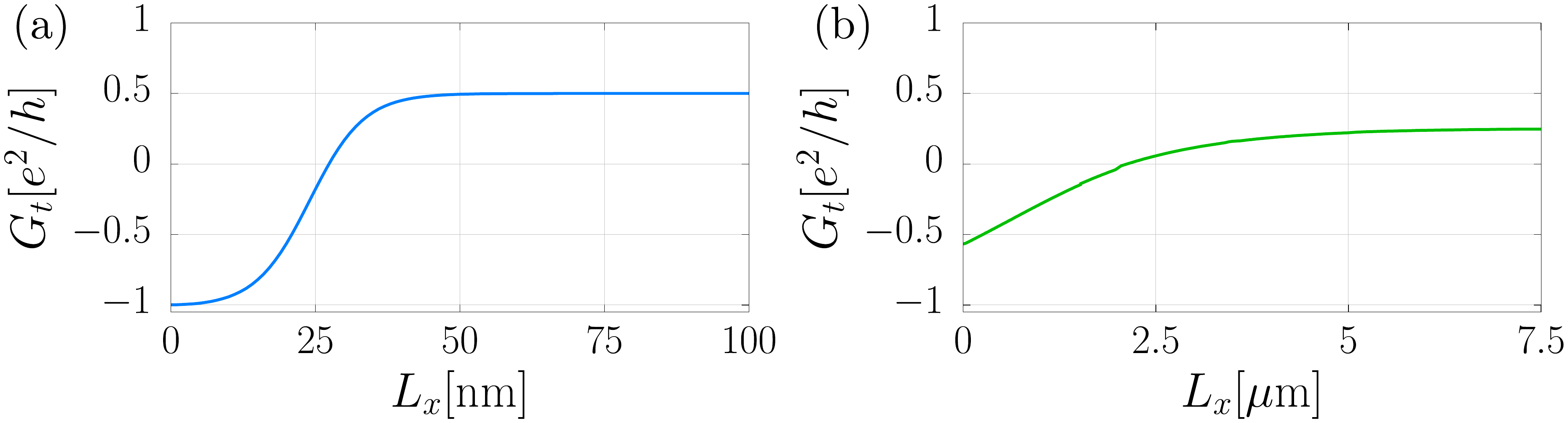}
    \caption{\label{fig:Barrier}
    Total conductance $G_t$ for a NSN'N junction with a proximitized sector in the (a) $N_{BDI}=1$ phase and (b) $\mathcal{N}=1$ TSC with an insulating barrier on the right side of the system. The conductance is computed as a function of the length $L_x$ of the barrier. The plots are obtained with $\alpha=0.25$.
    }
\end{figure}

A numerical simulation for the conductance in the NSN'N junction with $\alpha=0.25$ is shown in Fig.~\ref{fig:Barrier} for (a) a wire geometry with a proximitized sector in the $N_{BDI}=1$ state and (b) a film geometry with a $\mathcal{N}=1$ TSC in the central sector. We focus on the dependence
on $L_x$, the length of the intermediate barrier N'.
For a completely transparent barrier ($L_x \approx 0$) the total conductance
for $\alpha=0.25$ is $G_t=-e^2/h$ in presence of MBS and $G_t=-e^2/2h$ in presence of MCPS.
An increasingly opaque barrier ($L_x\to\infty$) changes these values, keeping $G_t \neq 0$ as long as Andreev processes occur in the proximitized MTI.
For the case of MBSs, a barrier with $L_x \gtrsim 50\, {\rm nm}$ is long enough to prevent the electric transmission on the right side, leading to $G_t=e^2/2h$.
Remarkably, Fig.~\ref{fig:Barrier}(b) shows a very different decay length along $x$ for MCPSs.
Indeed, a larger barrier $L_x \gtrsim 5\; \mu{\rm m}$ is required to prevent completely the electric transmission in presence of MCPSs, changing the total conductance to $G_t=e^2/4h$.
Both limiting values are in  agreement with Table~\ref{tab:G_barrier} for the selected bias split parameter $\alpha=0.25$.
Focusing on the short barrier limit, which may represent interface disorder effects, Fig.~\ref{fig:Barrier} suggests that, for the chosen set of parameters, the antisymmetry of $G_t(\alpha)$ is robust for barriers with $L_x\lesssim 5$ nm for the MBS and $L_x\lesssim 0.5$ $\mu$m for the MCPS, since $G_t$ is almost unaffected by the barrier in these cases.

\bibliography{bibliography}

\end{document}